\pdfoutput=1

\documentclass[letterpaper, 10 pt, conference]{ieeeconf}  

\IEEEoverridecommandlockouts                              
\usepackage{graphics} 
\usepackage{epsfig} 
\usepackage{mathptmx} 
\usepackage{times} 
\usepackage{amsmath} 
\usepackage{amssymb}  
\usepackage{textcomp}
\usepackage{amsmath}
\let\labelindent\relax
\usepackage{enumitem}

\DeclareMathOperator*{\argmax}{arg\,max}
\title{\LARGE \bf HeartBEAT: Heart Beat Estimation through Adaptive Tracking
}

\author{Huijie Pan, Dogancan Temel and Ghassan AlRegib
\thanks{Center for Signal and Information Processing (CSIP), School of Electrical and Computer Engineering,
        Georgia Institute of Technology, Atlanta, GA 30332-0250, USA}%
        }
        
\begin{document}

\twocolumn[{%

\vspace{40mm}

\begin{itemize}[leftmargin=2.5cm, align=parleft, labelsep=2.0cm, itemsep=4ex,]

\item[\textbf{Citation}]{H. Pan, D. Temel and G. AlRegib, "HeartBEAT: Heart beat estimation through adaptive tracking," 2016 IEEE-EMBS International Conference on Biomedical and Health Informatics (BHI), Las Vegas, NV, 2016, pp. 587-590.
}

\item[\textbf{DOI}]{https://doi.org/10.1109/BHI.2016.7455966}

\item[\textbf{Online}]{https://ghassanalregib.com/publications/}

\item[\textbf{Bib}] {@ARTICLE\{Temel2016\_BHI,\\ 
author=\{H. Pan and D. Temel and G. AlRegib\},\\ 
booktitle=\{2016 IEEE-EMBS International Conference on Biomedical and Health Informatics (BHI)\},\\ 
title=\{HeartBEAT: Heart beat estimation through adaptive tracking\}, \\ 
year=\{2016\},\\ 
pages=\{587-590\},\\ 
doi=\{10.1109/BHI.2016.7455966\},\\ 
ISSN=\{2168-2208\},\\ 
month=\{Feb\},\}
} 

\item[\textbf{Copyright}]{\textcopyright 2016 IEEE. Personal use of this material is permitted. Permission from IEEE must be obtained for all other uses, in any current or future media, including reprinting/republishing this material for advertising or promotional purposes,
creating new collective works, for resale or redistribution to servers or lists, or reuse of any copyrighted component
of this work in other works. }

\item[\textbf{Contact}]{alregib@gatech.edu~~~~~~~https://ghassanalregib.com/ \\ dcantemel@gmail.com~~~~~~~http://cantemel.com/}
\end{itemize}

}]

\maketitle
\thispagestyle{empty}
\pagestyle{empty}

\begin{abstract}

In this paper, we propose an algorithm denoted as \texttt{HeartBEAT} that tracks heart rate from wrist-type photoplethysmography (PPG) signals and simultaneously recorded three-axis acceleration data.  \texttt{HeartBEAT} contains three major parts: spectrum estimation of PPG signals and acceleration data, elimination of motion artifacts in PPG signals using recursive least Square (RLS) adaptive filters, and auxiliary heuristics. We tested  \texttt{HeartBEAT} on the 22 datasets provided in the 2015 IEEE Signal Processing Cup. The first ten datasets were recorded from subjects performing forearm and upper-arm exercises, jumping, or pushing-up. The last twelve datasets were recorded from subjects running on tread mills. The experimental results were compared to the ground truth heart rate, which comes from simultaneously recorded electrocardiogram (ECG) signals. Compared to state-of-the-art algorithms, \texttt{HeartBEAT} not only produces comparable Pearson's correlation and mean absolute error, but also higher Spearman\textquotesingle s $\rho$ and Kendall\textquotesingle s $\tau$.

\end{abstract}

\section{Introduction}

Heart rate monitoring can provide useful information for people who are in rehabilitation or in pysical exercises and anyone who wants to routinely keep track of their vital signs. Ever since their emergence, wearable devices have become more and more popular for heart rate monitoring. Most of these wearable devices use PPG signal sampled on a subject\textquotesingle s ear, fingertip or wrist to track the subject\textquotesingle s heart rate. Unfortunately, PPG signals are easily contaminated by motion artifacts, interference caused by a subject\textquotesingle s motion \cite{c1}. Therefore, to track a subject\textquotesingle s heart rate by analyzing the sampled PPG signal, methods have to be introduced to circumvent the motion artifacts.

Several methods have been proposed in the literature for heart rate tracking. Kim and Yoo \cite{c2} use independent component analysis (ICA) to reduce motion artifacts in PPG signals, assuming that motion artifacts and uncorrupted PPG signals are statistically independent. However, Yao and Warren \cite{c3} have shown that such an independence is unwarranted, and cautions should be made when one uses ICA to eliminate motion artifacts. Others propose using singular value decomposition \cite{c4}, time and period domain analysis \cite{c5}, wavelet transform \cite{c6}, Wigner-Ville distribution \cite{c7}, and pure heuristics \cite{c8} to track heart rate. The shortcoming of most of the state of the art approaches in the literature is that they can only operate under limited motion such as running on treadmill or minor finger movements.

In this paper, we propose  \texttt{HeartBEAT}  that tracks a subject\textquotesingle s heart rate from the PPG signal that is subject to various motion artifacts. In Section \ref{sec:algorithm}, we describe the main blocks in the algorithm where spectrum estimation using periodogram is explained in Section \ref{subsec:spectrum}, RLS adaptive filtering is explained in Section \ref{subsec:adaptive}, and heuristics is explained in Section \ref{subsec:heuristics}. We describe the datasets in Section \ref{sec:data}, discuss the results in Section \ref{sec:results}, and conclude the work in Section \ref{sec:conclusions}. 

\section{Algorithm Details}
\label{sec:algorithm}
\begin{figure*}[thpb]
      \centering

{\includegraphics[width=0.7\linewidth]{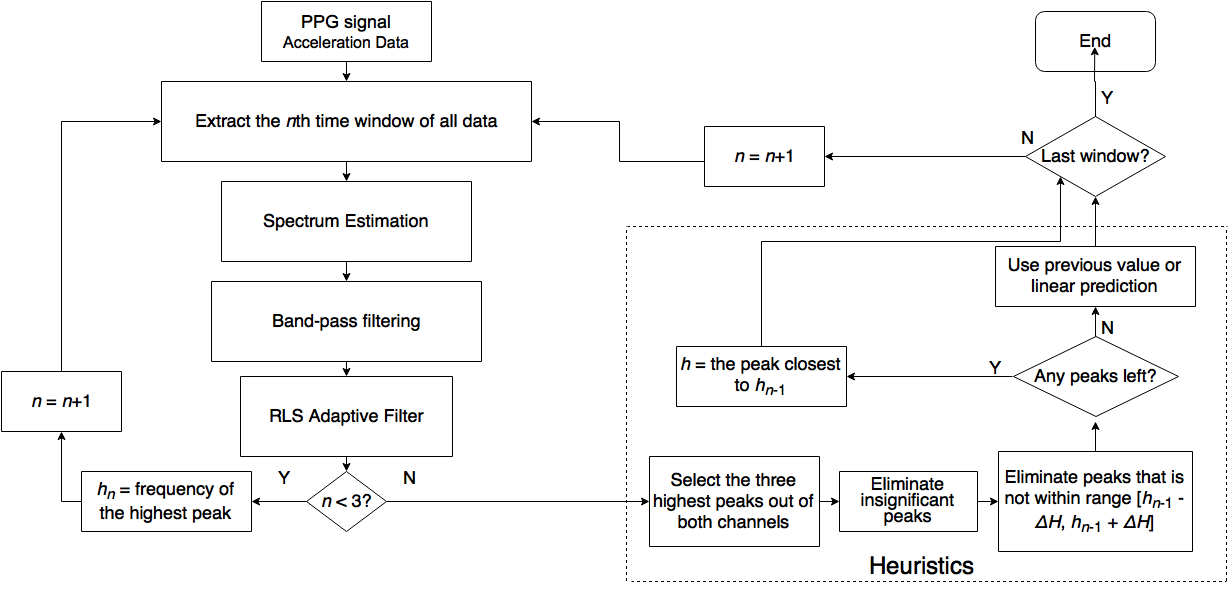}}
      \caption{Flow chart of \texttt{HeartBEAT} .}
      \label{figurelabel1}
      \vspace{-5mm}
   \end{figure*}

The overall structure of  \texttt{HeartBEAT}  is shown in Fig.~\ref{figurelabel1}. The algorithm is specifically designed for PPG data with simultaneously recorded three-axis acceleration. In our case, we have two channels of PPG data for each subject. The two channels of PPG came from two pulse oximeters emitting green lights. The distance between them was $2$ cm. The acceleration data came from a tri-axis accelerometer (Refer to section \ref{sec:data} for more description of the data sets). In the following subsections, we explain each step of our algorithm in details.

\subsection{Spectrum Estimation using Periodogram}
\label{subsec:spectrum}
First, let us show the low resolution aspect of the data sets. The sampling rate of the PPG signal and the acceleration data is 125 samples/sec, and the normal range of frequency of the heart rate is $\frac { 2 }{ 3 }$ Hz to $4$ Hz. Therefore, the portion of the spectra of the PPG signal and the acceleration data in which we are interested given by FFT will have a low resolution. In the proposed work, we would like to analyze the portion of the spectra in the normal human heart rate range, which is from $40$ beats/min to $240$ beats/min. If we analyze all the signal based on $8$-second windows with $6$-second overlaps, each window will have 1000 data points. The spectrum of the data points in each window by FFT has the same length. These 1000 data points are evenly distributed in the normalized frequency range, $0-2\pi$, which corresponds to $0$ - $125$ Hz. In the range of interest, $\frac { 2 }{ 3 }$ Hz to $4$ Hz, only 27 data points are available. Adjacent two data points have a $0.125$ Hz (= $7.5$ beats/min) difference in frequency, which leads to significant loss of information in the spectra. 
To solve this problem, we use spectrum estimation, which provides a trade off between resolution and accuracy. The specific method we are using is periodogram \cite{c11}. The periodogram of a signal, ${\{{x}_{k}\} (k=1, 2, ..., N)}$, is given in Eq. $1$

$$
S(f) = \frac { { f }_{ s } }{ N } { \left| \sum _{ k=1 }^{ N }{ { x }_{ k }{ e }^{ -j2\pi kf } }  \right|  }^{ 2 } \eqno{(1)}
$$ where ${ { f }_{ s } }$ is the sampling frequency, ${N}$ is the length of ${\{{x}_{k}\}}$, and ${S(f)}$ is the spectral density.

\begin{figure}[thpb]
      \centering
      {\includegraphics[width=0.75\linewidth] {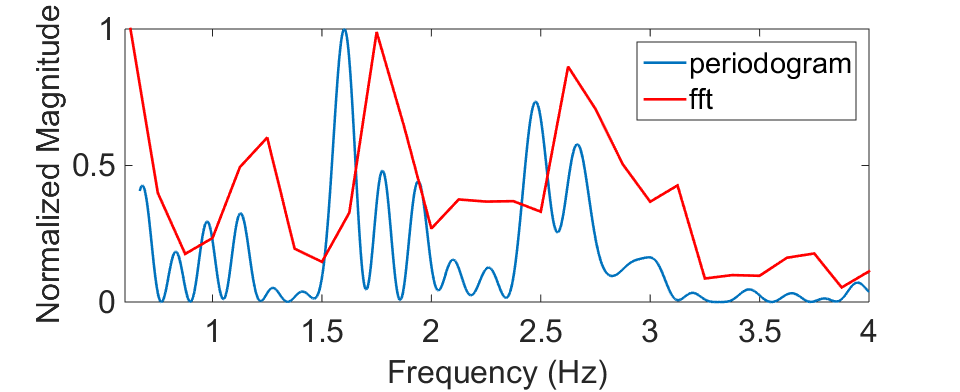}}
      \caption{Comparison of the FFT and the periodogram of a specific time window of PPG signal.}
      \label{figurelabel2}
        \vspace{-2mm}
   \end{figure}

Fig.~\ref{figurelabel2} shows a comparison between the spectrum of a window of PPG signal by FFT and that by periodogram. Both have been band-pass filtered. The cutoff frequencies are $\frac { 2 }{ 3 }$ Hz and $4$ Hz, or $40$ beats/min and $240$ beats/min. As it can be observed in Fig.~\ref{figurelabel2}, the periodogram has a much higher resolution and the low resolution of FFT results in loss of spectral information.
	
\subsection{RLS Adaptive Filtering}
\label{subsec:adaptive}
\begin{figure}[thpb]
      \centering
      {\includegraphics[width=0.75\linewidth] {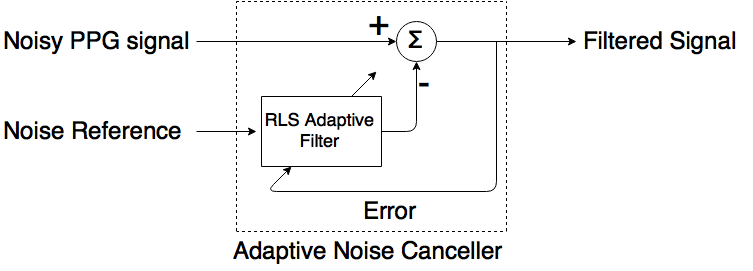}}
      \caption{Structure of an RLS adaptive noise canceler.}
      \label{figurelabel3}
        \vspace{-4mm}
   \end{figure}

Intuitively, a fixed-frequency motion should introduce a signal with the same frequency in the PPG spectrum. Based on this intuition, a peak with significantly high magnitude that appears in the spectrum of acceleration results in a peak  appearing at the same frequency in the spectra of both PPG channels. In some cases, acceleration-based peaks in the PPG spectra are more significant compared to the peaks near the ground truth heart rate. Therefore, we consider the acceleration as a noise reference for part of the motion artifacts in the PPG signal. RLS adaptive filters, which can minimize the error in the input signal in the least square sense given the noise reference, are successfully used in the literature to denoise signals when the noise reference is available \cite{c12}. Fig.~\ref{figurelabel3} shows the general structure of an RLS adaptive noise canceling filter. For our datasets, we used a 16-length householder RLS filter. Longer length increases the computational cost, while shorter length is not as effective in eliminating noise.

\begin{figure}[t]
      \centering
      \begin{minipage}[b]{0.6\linewidth}
      \centering
      \centerline{\includegraphics[width=1\linewidth]{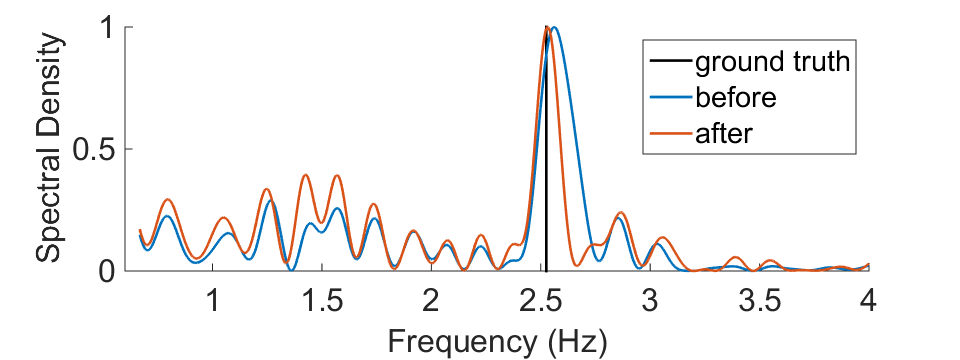}}
      \centerline{(a) Channel 1 PPG}\medskip
      \end{minipage}
      \begin{minipage}[b]{0.6\linewidth}
      \centering
      \centerline{\includegraphics[width=1\linewidth]{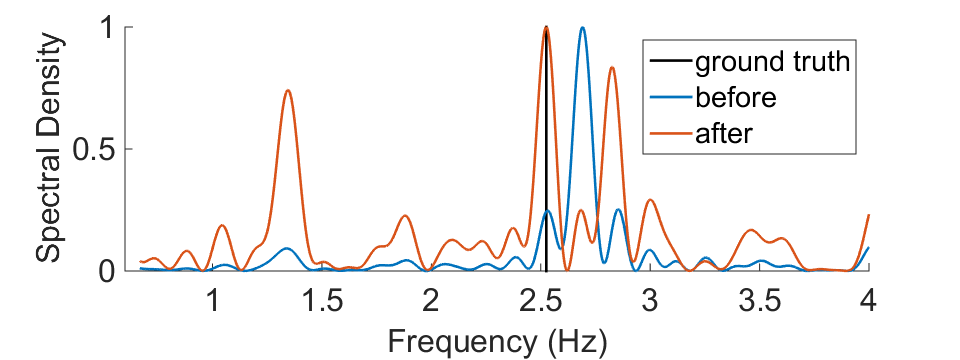}}
      \centerline{(b) Channel 2 PPG}\medskip
      \end{minipage}
      \begin{minipage}[b]{0.6\linewidth}
      \centering
      \centerline{\includegraphics[width=1\linewidth]{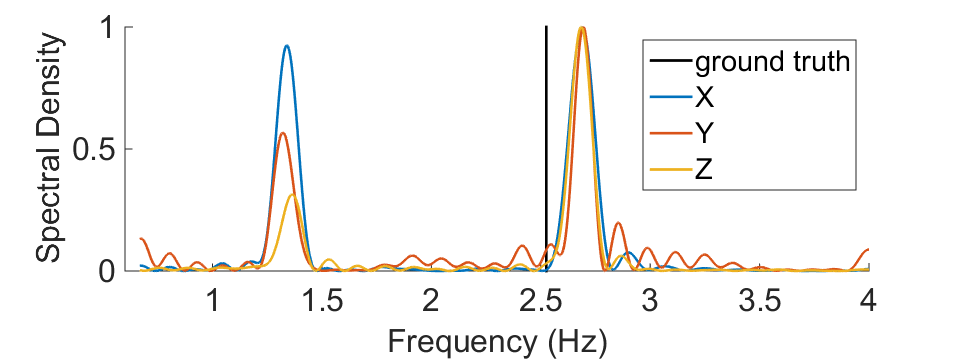}}
      \centerline{(c) Acceleration}\medskip
      \end{minipage}
      \vspace{-2mm}
      \caption{Three plots showing the effectiveness of the RLS adaptive filter.}
      \label{figurelabel4}
      \vspace{-2mm}
   \end{figure}

 Spectra of a few PPG signals are given in Fig.~\ref{figurelabel4}. The first PPG channel is given in Fig.~\ref{figurelabel4}(a) and the second PPG channel in Fig.~\ref{figurelabel4}(b). These spectras show the waveform before and after the adaptive filter from one data set at a specific time. Fig.~\ref{figurelabel4}(c) shows the spectrum of the tri-axis acceleration that corresponds to the same time interval. It can be easily seen that, after the adaptive filtering stage, in the spectra of both channels of PPG signal, peaks near the ground truth heart rate remain similar whereas motion-based peaks are attenuated, and become less significant than those near ground truth heart rates, which articulates the effect of the RLS filter.

\subsection{Heuristics}
\label{subsec:heuristics}
\begin{figure}[t]
      \centering
      \begin{minipage}[b]{0.6\linewidth}
      \centering
      \centerline{\includegraphics[width=1\linewidth]{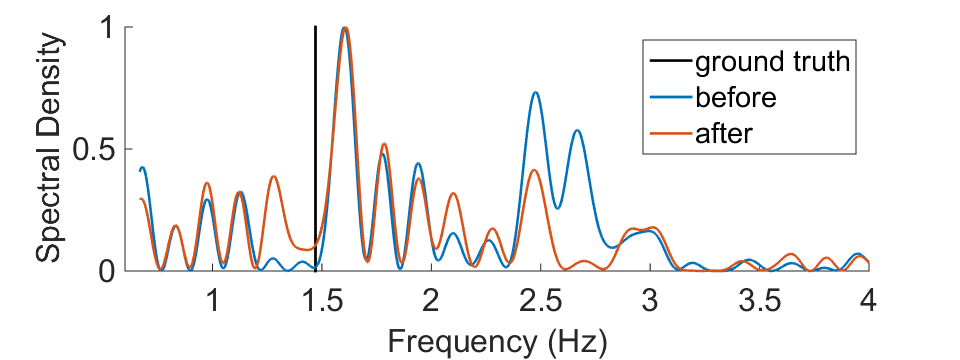}}
      \centerline{(a) Channel 1 PPG}\medskip
      \end{minipage}
      \begin{minipage}[b]{0.6\linewidth}
      \centering
      \centerline{\includegraphics[width=1\linewidth]{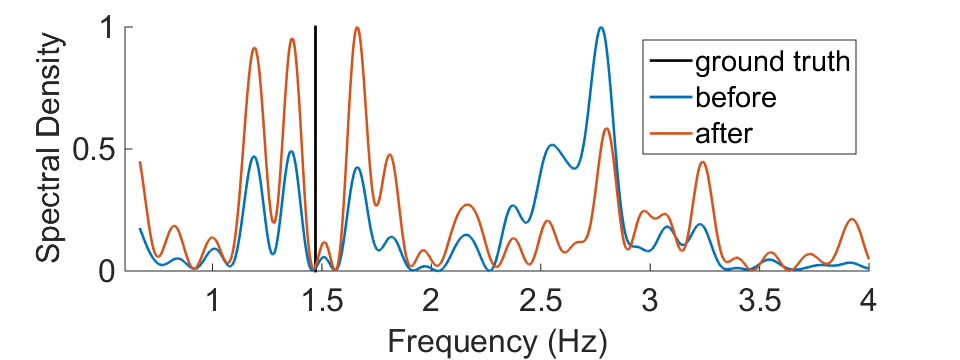}}
      \centerline{(b) Channel 2 PPG}\medskip
      \end{minipage}
      \begin{minipage}[b]{0.6\linewidth}
      \centering
      \centerline{\includegraphics[width=1\linewidth]{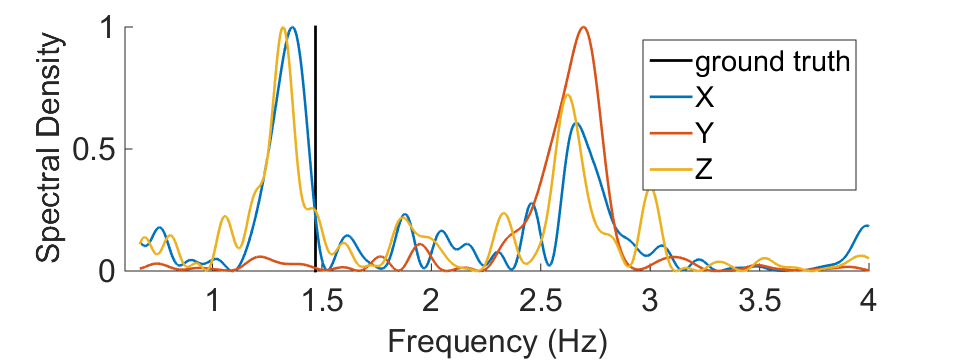}}
      \centerline{(c) Acceleration}\medskip
      \end{minipage}
      \caption{Three plots showing case where the RLS adaptive filter does not work.}
      \label{figurelabel5}
      \vspace{-6mm}
   \end{figure}

Even though the RLS adaptive filter is able to eliminate false peaks in the PPG spectra, it fails under various circumstances. We are not able to generalize the situations where the adaptive filter fails. One possible explanation is that at some windows the noise in the PPG signals may come from other sources other than motion. One example is shown in Fig.~\ref{figurelabel5}. Unlike the case in Fig.~\ref{figurelabel4}, in this case the adaptive filter does not eliminate the peaks in the PPG spectra related to the ones in the acceleration spectrum. When the adaptive filter fails, we can use the heuristics that are based on the following observations.

Initially, the subject is at rest, and the highest peak in the PPG spectrum is the heart rate. Let ${ h }_{ n }$ be the heart rate selected by the algorithm and ${S(f)}$ be the spectrum of channel 1 PPG in the current time window, then
$$
{ h }_{ n }=\underset { f }{ \argmax S(f) },\quad n\le 3  \eqno{(2)}
$$After the first 3 time windows when the highest peak is assumed to be the heart rate, the algorithm will select peaks within a certain range with respect to the previous heart rate. We assume the maximum allowed change of heart rate from one time window to the next to be $\Delta H$ (which stands for heart rate change limit and whose initial value is 12 beats/min), then the algorithm will first find peaks in the spectra of both channels of the PPG signal, and determine if any of them is in the range $\left[ { h }_{ n-1 }-\Delta H,\quad { h }_{ n-1 }+\Delta H \right]$. If there is only one peak in this range, then this peak will be selected. If there are more than one valid peaks, then the highest will be selected. If there are two valid peaks with the same magnitude, then the one closer to ${ h }_{ n-1 }$ will be selected.
If there is no valid peak, then the previous heart rate value will be used. Namely,
$$
{ h }_{ n }={ h }_{ n-1 }.\eqno{(3)}
$$
Each time the previous heart rate is used, $\Delta H$ is increased by 5 in the next round. $\Delta H$ is increased more if the previous heart rate is still used to prevent error propagation. Once the algorithm detects a peak, $\Delta H$ is reset to its initial value. To prevent continuous error propagation, using the previous heart rate is only allowed four times in a row. If after four consecutive times of using previous heart rate, still no valid peaks exist, a prediction will be made based on the previous heart rates. To make a prediction, the algorithm will first extrapolate a point based on the linear regression of its previous 10 points (or all previous points, if there are less than 10 points). Here 10 is used because for a larger value, the heart rate trend within the previous time windows on which the prediction is based is more likely to be nonlinear, and a smaller value reduces the reliability of the linear regression. Let the linearly extrapolated point be ${ h }_{ e }$, then
$$
{ h }_{ n }={ h }_{ n-1 }+5*sign({ h }_{ e }-{ h }_{ n-1 }). \eqno{(4)}
$$ where $sign(x)$ denotes the signum of ${x}$. The multiplier of the sign function is chosen based on experimenting with subset of the datasets. After the prediction method is used, $\Delta H$ is increased by 8.

\section{Data Set Description}
\label{sec:data}
The 22 data sets are introduced as part of the 2015 IEEE Signal Processing Cup and used by various researchers \cite{c1}\cite{c9}\cite{c10}. All sampling rates were $125$ samples/s. The data were recorded from subjects performing different kinds of motions. Specifically, data sets 1, 2, 7, and 10 were recorded from subjects performing various forearm and upper arm exercise (like shaking hands, stretching and pushing), running, jump, and push-up. Data sets 3 - 6 and 8 - 9 were recorded from subjects performing intensive arm movements like boxing; data set 11 was recorded from a subject running on a tread mill: the speed was rest($30$ s) $\rightarrow$ $8$ km/h($1$ min) $\rightarrow$ $15$ km/h($1$ min) $\rightarrow$ $8$ km/h($1$ min) $\rightarrow$ $15$ km/h($1$ min) $\rightarrow$ rest($30$ s); data sets $12$ - $22$ were recorded from subject running on tread mills: their speed was rest($30$ s) $\rightarrow$ $6$ km/h($1$ min) $\rightarrow$ $12$ km/h($1$ min) $\rightarrow$ $6$ km/h($1$ min) $\rightarrow$ $12$ km/h($1$ min) $\rightarrow$ rest($30$ s). Data set 10 was recorded from a subject with abnormal heart rhythm and blood pressure, and thus the estimation result of this algorithm may only reflect the subject's pulse rate. Data sets 1 - 9 come from healthy subjects. Information regarding the health of subjects in data sets 10 - 22 is unavailable from the Signal Processing Cup.

\section{Results}
\label{sec:results}
To measure the performance of \texttt{HeartBEAT}, four metrics were used: Pearson\textquotesingle s correlation, Spearman\textquotesingle s $\rho$, Kendall\textquotesingle s $\tau$ and mean absolute error with respect to the ground truth. Let the tracked heart rate be ${{ h }_{ i }}$, the ground truth heart rate be ${{ G }_{ i }}$, and the total number of heart rates be ${n}$. Then, the Pearson's Correlation is given by
$$
{ \rho  }_{ h,G }=\frac { cov(h,G) }{ { \sigma  }_{ h }{ \sigma  }_{ G } } \eqno{(5)}
$$ where ${ cov(h,G) }$ is the covariance of ${h}$ and ${G}$, and ${ \sigma  }$ is the standard deviation. 
The Spearman\textquotesingle s $\rho$ is given by
$$
\rho =1-\frac { 6\sum { { d }_{ i }^{ 2 } }  }{ n({ n }^{ 2 }-1) } \eqno{(6)}
$$ where ${ { d }_{ i } } $ is the difference of ranks of ${{ h }_{ i }}$ and ${{ G }_{ i }}$. The Kendall\textquotesingle s $\tau$ is given by
$$
\tau =\frac { { n }_{ c }-{ n }_{ d } }{ \frac { 1 }{ 2 } n(n-1) } \eqno{(7)}
$$ where ${{ n }_{ c }}$ is the number of concordant pairs and ${{ n }_{ d }}$ is the number of discordant pairs. The mean absolute error is given by
$$
MAE = \frac { 1 }{ n } \sum _{ i=1 }^{ n }{ \left| { h }_{ i }-{ G }_{ i } \right|.  } \eqno{(8)}
$$


The 22 sets of data were also used in other papers \cite{c1}\cite{c9}. Table ~\ref{table1} compares the results of the proposed algorithm and the state of the art on these 22 sets. Another paper \cite{c10} uses only the last 12 data sets, recorded from subjects running on treadmills. Table ~\ref{table2} compares the four algorithms on the last 12 sets of data. 
For the 22 data sets, it can be seen that \texttt{HeartBEAT} has comparable Pearson's correlation and mean absolute error, higher Spearman\textquotesingle s $\rho$ and higher or equivalent Kendall\textquotesingle s $\tau$. \texttt{HeartBEAT}'s performance is better or equivalent in terms of Spearman and Kendall which means that the monotonic behavior is as good as or outperforming other methods but it still needs enhancement in terms of linear behavior which can be achieved by adding a monotonic mapping function. For running-on-treadmill motions (data sets 11-22), \texttt{HeartBEAT} produces not as good but comparable results. 

\begin{table}[!htbp]
\caption{Algorithm Result Comparison on Data Set 1-22}
\label{table1}
\begin{center}
\begin{tabular}{|c|c|c|c|c|}
\hline
Reference & Pearson & Spearman & Kendall & MAE\\
\hline
\cite{c1} & 0.941 & 0.913 & 0.823 & 2.62\\
\hline
\cite{c9}& 0.958 & 0.930 & 0.860 & 2.09\\
\hline
\texttt{HeartBEAT} & 0.932 & 0.935 & 0.860 & 2.19\\
\hline
\end{tabular}
\end{center}
\end{table}

\begin{table}[!htbp]
\caption{Algorithm Result Comparison on Data Set 11-22}
\label{table2}
\begin{center}
\begin{tabular}{|c|c|c|c|c|}
\hline
Reference & Pearson & Spearman & Kendall & MAE\\
\hline
\cite{c1} & 0.992 & 0.976 & 0.906 & 2.34\\
\hline
\cite{c9}& 0.993 & 0.988 & 0.943 & 1.28\\
\hline
\cite{c10} & 0.989 & N/A & N/A & 1.83\\
\hline
\texttt{HeartBEAT} & 0.982 & 0.984 & 0.928 & 1.56\\
\hline
\end{tabular}
\end{center}
\end{table}

\section{Conclusions and Future Work}
\label{sec:conclusions}
In this paper, we proposed \texttt{HeartBEAT} algorithm for tracking heart rate from PPG signal with various types of motion artifacts, including forearm and upper arm movements, push-up, jump and running on treadmills. \texttt{HeartBEAT} contains spectrum estimation using periodogram, motion artifacts elimination by recursive least square adaptive filters, and heuristics. Twenty-two sets of data were used to test the algorithm. The results produced by \texttt{HeartBEAT} shows high correlation and low mean absolute error with respect to the ground truth. Compared to other algorithms, it has lower correlation in terms of linearity. However, in terms of ranking-based methods, it is either the best or the second best. As future work, we are figuring out what kind of characteristics related to running-on-treadmill motions cause \texttt{HeartBEAT} to perform poorly. We are also enhancing the algorithm to make it work for various signals types other than PPG and also to make it robust under more significant motion artifacts so that \texttt{HeartBEAT} can potentially be used on wearable devices.

\section{ACKNOWLEDGMENT}
The authors of this paper would like to thank Dr. Zhilin Zhang, who is with Samsung Research America - Dallas, for providing the results of his papers. The authors would also like to thank Yuting Hu and Zhen Wang, PhD students at the Multimedia Sensors Lab at the Georgia Institute of Technology for their technical discussions and feedback on this paper.
\addtolength{\textheight}{-12cm}   






\end{document}